\documentclass[12pt,preprint]{aastex}
\usepackage{epsfig}
\usepackage{natbib}

\newcommand\kms{\:\rm{\,km\,s^{-1}}}

\newcommand\masy{\:\rm{\,mas\:yr^{-1}}}

\newcommand{\rx}{RX J0822$-$4300}
\newcommand{\pu}{Puppis A}

\shorttitle{{The proper motion of \rx}}
\shortauthors{Werner Becker, Tobias Prinz, P.~Frank Winkler, Robert Petre}

\begin{document}
\title{{The Proper Motion of the Central Compact Object \linebreak \rx\ in the Supernova Remnant \pu}}
\author{ Werner Becker\altaffilmark{1,2}, Tobias Prinz\altaffilmark{1}, P.~Frank Winkler\altaffilmark{3}, 
Robert Petre\altaffilmark{4}}

\altaffiltext{1}{Max-Planck Institut f\"ur extraterrestrische Physik, Giessenbachstrasse 1, 85741 Garching, Germany}
\altaffiltext{2}{Max-Planck Institut f\"ur Radioastronomie, Auf dem H\"ugel 69, 53121 Bonn, Germany}
\altaffiltext{3}{Department of Physics, Middlebury College, Middlebury, VT 05753}
\altaffiltext{4}{NASA Goddard Space Flight Center, Greenbelt, MD 20771}

\begin{abstract}
\noindent
 Using the High Resolution Camera (HRC) aboard the {\em Chandra} X-ray Observatory, we have re-examined 
 the proper motion of the central compact object \rx\, in the supernova remnant Puppis A. New data
 from 2010 August, combined with three  archival data sets from as early as 1999 December, provide 
 a baseline of 3886 days (more than 10\,\,1/2 years) to perform the measurement. Correlating the 
 four positions of \rx\, measured in each data set implies a projected proper motion of 
 $\mu= 71 \pm 12 \masy$. For a distance of 2 kpc this proper motion is equivalent to a 
 recoil velocity of $672 \pm 115 \kms $. The position angle is found to be $244\pm 11$ degrees. 
 Both the magnitude and direction of the proper motion are in agreement with \rx\, originating
 near the optical expansion center of the supernova remnant. For a displacement of $371 \pm 31$ 
 arcsec between its birth place and today's position we deduce an age of $(5.2\pm
 1.0)\, 10^3$ yrs for \rx. The age inferred from the neutron star proper motion and filament 
 motions can be considered as two independent measurements of the same quantity. They average 
 to $4450 \pm 750$ yrs for the age of the supernova remnant Puppis A.
\end{abstract}


\section{Introduction \label{intro}}

 There has long been broad consensus that core-collapse supernovae---the explosions of massive 
 progenitors that produce Types II, Ib, and Ic events at least---leave behind a compact stellar 
 remnant: either a neutron star or a black hole.  Numerous central compact objects (CCOs) have 
 been found within supernova remnant (SNR) shells (cf.~Becker 2010 for a review), including near 
 the centers of all three of the  ``oxygen-rich" SNRs (those with optical filaments dominated 
 by ejecta that stem only from core-collapse supernovae) in the Galaxy: Cas A \citep{tananbaum99, 
 chakrabarty01}, Puppis A \citep{petre96}, and  G292.0+1.8 \citep{hughes01, camilo02}.  

 Observations of proper motions for these CCOs, and of the ejecta-dominated filaments, provide 
 the opportunity to investigate the dynamics of core-collapse explosions, as imprinted on the 
 compact remnant and the ejecta.  Anisotropies and/or bipolar jets appear typical of the ejecta 
 distribution from core-collapse explosions.  These are predicted from two- and three-dimensional 
 models for core-collapse SNe \citep[e.g.,][]{burrows95, scheck04, scheck06, wong10}, and have been 
 observed in all three of 
 the Galactic oxygen-rich remnants.  Asymmetries in the ejecta distribution are closely linked 
 to the recoil of a CCO through conservation of momentum: if the explosion expels ejecta 
 preferentially in one direction, the CCO must recoil in the opposite direction.  One expects 
 this recoil to be particularly apparent in Puppis A, where \citet{Winkler88} found that all 
 the visible ejecta knots are moving generally toward the north and east.

 Following the discovery of \rx\ near the center of Puppis A \citep{petre96}, measuring 
 its proper motion and recoil velocity was of obvious interest, and observations with this 
 goal were carried out early in the {\em Chandra} mission.  We have previously done two independent 
 studies of  the proper motion of \rx\ in a pair of papers:  \citet{HuiBeck06}, henceforth HB06, 
 and \citet{Winkler07}, henceforth WP07. Both these papers were based on the same three epochs 
 of data from the {\em Chandra} HRI over the 5.3-year period 1999 December--2005 April, and both measured 
 a motion to the southwest, as expected, but with surprisingly high velocity:  HB06 measured 
 $\mu = 107\pm 34 \masy$, while WP07 found $165 \pm 25 \masy$, corresponding to a transverse 
 velocity of $1122 \pm 327 \kms$, or $1570 \pm 240 \kms$, respectively, if we take Puppis A's 
 distance as 2 kpc. Both these values are large compared with the typical birth velocities for  
 pulsars of $\sim 500 \kms$ \citep{caraveo93, frail94, hobbs05}.  Furthermore, a recoil velocity 
 much larger than $1000 \kms$  challenges  models for producing pulsar kicks.  Therefore, we have 
 undertaken this joint follow-up study, taking advantage of {\em Chandra}'s unique spatial resolution 
 with a  time baseline twice as long as for our previous measurements.

\section{Observations and Data Reduction\label{obs}}
 During the course of the {\em Chandra} mission, \rx~has been observed on four occasions with the High 
 Resolution Camera (HRC), the first of these in 1999 December.  The first three observations have 
 been already archived for several years.  Because of the importance of \rx\ and the somewhat 
 discrepant results found by HB06 and WP07, a new observation was proposed and carried out 2010 
 August 10-11. The new observation was carried out as two consecutive ObsIDs, with exposure 
 times of  $\sim 40$ and $\sim 38$ ks, with no intervening repointing, so they we have merged 
 the event data for these in our analysis, for a total exposure of 78.9 ks. All observations 
 are summarized in Table \ref{table1}.

 We  downloaded fresh copies of the archived observations, and have  reprocessed the events using 
 the {\tt chandra\_repro} script.  This and all subsequent analysis has been carried out using 
 CIAO 4.4, and CALDB 4.4.8 to ensure that the latest  corrections  have been applied uniformly.

\begin{figure}
\centerline{\psfig{figure=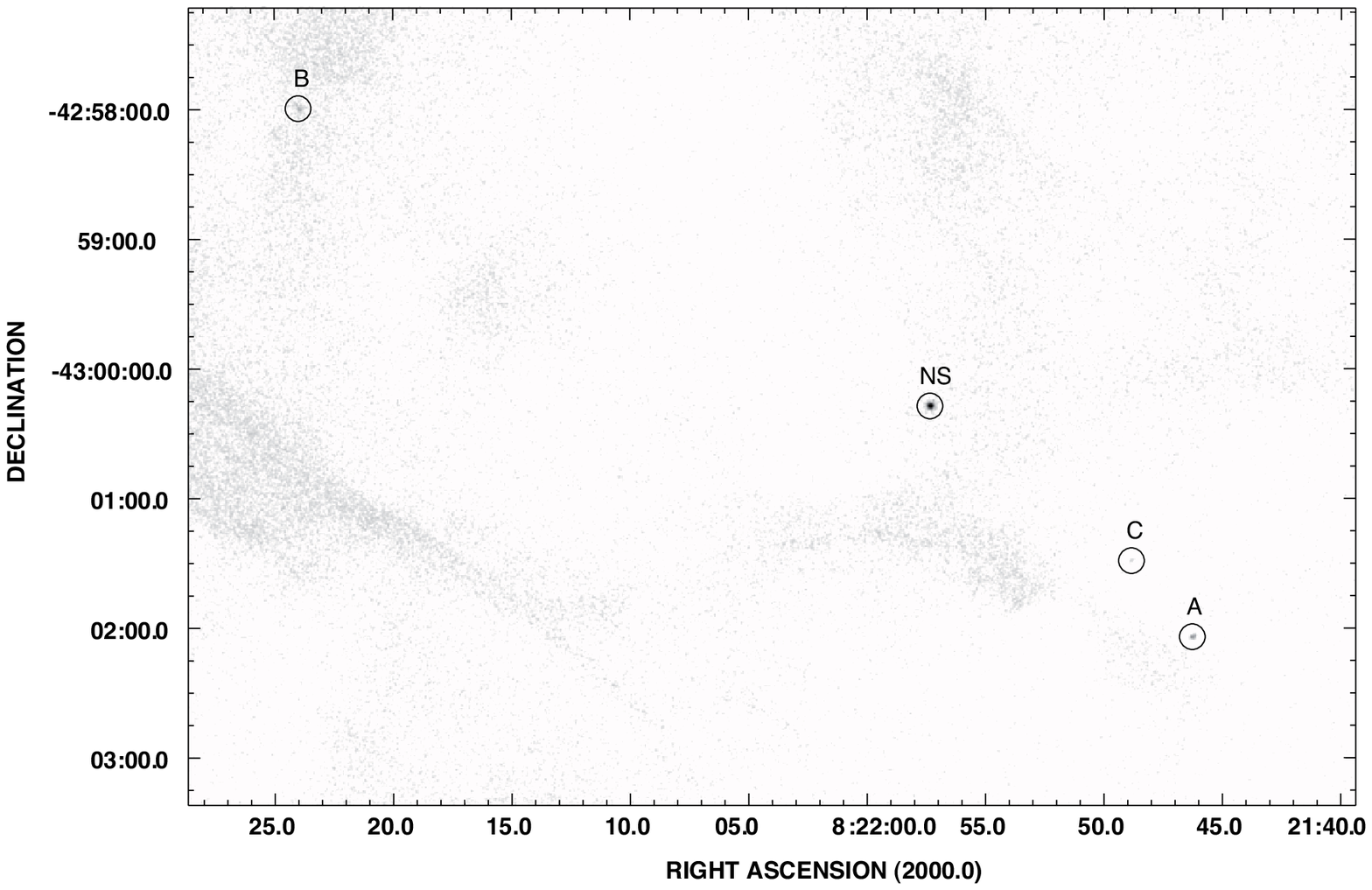,width=16cm,clip=}}
\caption[]{The 2010 epoch {\em Chandra} HRC-I image with reference sources and \rx~marked 
by circles. NS marks the position of \rx; A, B and C that of the fiducial stars 
used as local calibrators for absolute astrometry. The field measures 
6 by 9 arcmin and is oriented north up, east left.}
\label{figure1}
\end{figure}

 At the $\sim 2$ kpc distance of Puppis A, a $1000 \kms$ transverse velocity of \rx\ would result 
 in a proper motion of only $\sim 0.1''\,\mbox{yr}^{-1}$. Even for {\em Chandra}, whose absolute aspect 
 resolution\footnote{\url{http://cxc.harvard.edu/cal/ASPECT/celmon/}} is $\sim 0\farcs 6$, the 
 measurement of such a small proper motion is a challenge.  Fortunately, there are three nearby X-ray 
 sources  whose positions are very well determined, since all correspond to optical stars in the 
 UCAC3 \citep{Zacharias09} and 2MASS \citep{Cutri03} catalogs. Both HB06 and WP07 used these  
 stars for astrometric correction in their analysis of the $1999-2005$ data. We summarize their 
 properties in Table \ref{table2} using the same nomenclature as in HB06\@. All three sources 
 are detected with high significance in the 2010 observation. Using {\tt wavedetect}  we 
 determined their position and count rates, including those of \rx, as listed in Table 
 \ref{table3}. An image depicting their locations relative to \rx~is shown in Fig.~\ref{figure1}. 

 In order to achieve the highest accuracy in measuring the proper motion of RX J0822-4300,  we 
 followed the method described in \S3.4 of WP07  with some improvements: we first measured the X-ray 
 positions of \rx~and the reference stars A, B, and C in each of the three archival datasets 
 and in the 2010 observation  (\S \ref{spatial_analysis:1}). The X-ray and optical positions 
 of the stars were then used  to transform the respective images of each observation to an 
 absolute  reference frame (\S \ref{spatial_analysis:2}). This procedure is necessary in 
 order to include the HRC-S observation into our analysis, as there are small systematic 
 effects and offsets in the rotation between the HRC-I and HRC-S detectors (WP07). The 
 transformation coefficients determined this way were then used to derive the position of 
 the neutron star at the corresponding epoch. We calculate the  error budget exactly, rather 
 than use the approximate method of WP07, which leads to somewhat larger errors for the 
 corrected X-ray positions of \rx\ in the $1999-2005$ data when compared to their work.

\subsection{Spatial Analysis}\label{spatial_analysis:1}
 In order to determine the position of the three reference stars and \rx~we first generated 
 a point-spread function (PSF) for every source in each of the four observations. This is 
 required as the PSF broadens for sources at larger off-axis angles, leading to a larger 
 position uncertainty when using the {\tt wavedetect} tool. These positions were then used 
 as input to {\em ChaRT},\footnote{{\em Chandra} Ray Tracer, \url{http://cxc.harvard.edu/chart/index.html}} 
 which computed the PSF for a point source at any off-axis angle. The computations were performed 
 by generating 10 rays mm$^{-2}$ for a peak energy of either 1.5 keV (WB \& TP, the same as used by HB06) 
 or 1.0 keV  (FW, as in WP07);  the resulting PSFs are virtually identical. The rays were then projected 
 onto the HRC detector using {\em Marx}\footnote{\url{http://space.mit.edu/CXC/MARX/}} to create an image 
 of the PSF.

 We obtained best-fit positions for all three reference stars and for \rx\ at each of the epochs 
 using {\em Sherpa}, {\em Chandra}'s modeling and fitting package.\footnote{\url{http://cxc.harvard.edu/sherpa/}}   
 We followed the {\em Sherpa} thread ``Accounting for PSF Effects in 2D Image Fitting," in which a simulated 
 PSF is convolved with an assumed source model to produce the best fit to the actual observational data.  
 For each of the sources, we used a 2-D Gaussian model whose width was fixed at a small value -- in most 
 cases equal to the bin size used in the images -- so that small intrinsic fluctuations in different PSFs 
 and in the data did not lead to divergence. In our independent analyses, we used binnings of $0.5 \times 0.5$ 
 and $1 \times 1$ pixel for \rx, and $1 \times 1$ and $2 \times 2$ pixels for stars A, B, and C.  In each 
 case a PSF was constructed with {\em Marx} to match the bin size for the data.  Because of the relatively 
 low count rates for the three reference stars, we used the C-statistic to measure  the goodness of fit.  
 To obtain uncertainties, we used the {\em Sherpa} procedure {\tt reg\_proj} to map out 1$\sigma$, 
 90\%-confidence, and 2$\sigma$ contours in $(x, y)$ detector coordinates.  In every case, the best fits 
 obtained with different binning choices were consistent, but smaller bin sizes generally gave somewhat 
 smaller uncertainties.  

 Early in our data analysis we found a subtle software bug in the {\em Sherpa}  package for  CIAO 
 version 4.3. By default {\em Sherpa} was using the brightest pixel in the PSF-image as the reference 
 point for the convolution, regardless of the PSF-input-position specified to the task at the command 
 line. For near on-axis sources, the brightest pixel is always very close to the nominal PSF center, 
 but there can be significant displacements for a source that is a few arcminutes off-axis, with 
 direction differences that depend on roll angle.  This bug appears to be a long-standing one, and
 was apparently at work in the analysis reported in WP07, for two of us (WB \& TP) were able to 
 approximately reproduce those results for the $1999-2005$ data. (HB06 used a different approach 
 that did not involve PSF fits, so the result they reported was not affected.)

 The {\em Chandra} Help Desk was able to identify this bug and eventually provided us with a 
 workaround.\footnote{see \url{http://cxc.harvard.edu/sherpa/bugs/psf.html}}, which WB \& TP 
 have then used for all subsequent analysis.  This workaround has been incorporated into CIAO 
 version 4.4, which PFW has used to give essentially identical results.  There are other differences 
 between both the 2006-07 analyses and the present one that relate to the data themselves.  
 Subsequent to the earlier analyses, there has been a complete reprocessing of all the {\em Chandra} 
 data.  Differences include use of improved aspect files and updated values  for the degap 
 corrections, detector gain, and the telescope effective focal length. As noted earlier, the 
 present analysis has incorporated all the current wisdom regarding telescope and instrument 
 performance.  

 The fitted X-ray positions of \rx~(labeled as NS) and that of the three fiducial reference 
 stars are listed for each of the four observations in Table \ref{table3}, along the 
 respective HRC counting rates. 

\subsection{Transformation to the World Coordinate System (WCS)}\label{spatial_analysis:2}

 In order to determine the position of \rx~relative to the three reference stars we assume a 
 linear transformation with four free parameters: translations in Right Ascension, $t_{RA}$, 
 and in Declination, $t_{Decl}$, a scale factor $r$, and a rotation of the detector $\theta$. 
 The transformation can be expressed in the following way:
\begin{equation} \label{eq:transf}
\left( 
\begin{array}{cccc}
x_A & -y_A & 1 & 0 \\
y_A &  x_A & 0 & 1 \\
x_B & -y_B & 1 & 0 \\
y_B &  x_B & 0 & 1 \\
\end{array} 
\right) \left( \begin{array}{c}
r \cos\theta\\
r \sin\theta\\
t_{RA}\\
t_{Decl}\\
\end{array} \right) = \left( \begin{array}{c}
x'_A\\
y'_A\\
x'_B\\
y'_B\\
\end{array} \right),
\end{equation}
 where $x_i$, $y_i$ is the x-, and y-position of star $i$ in the HRC image at epoch $T$ and 
 $x'_A$, $y'_A$ are the corresponding optical coordinates of star $i$. These coordinates are 
 given by the UCAC3 catalog and are corrected for proper motion (see Table \ref{table2} and 
 Table \ref{table3}). We used stars A and B to calculate the transformation and star C to verify 
 the resulting parameters. Multiplying equation \ref{eq:transf} with the inverse of the matrix 
 leads to the missing parameters $t_x$, $t_y$, $r$ and $\theta$.  The position of \rx~at epoch 
 $T$ can then be calculated straightforwardly by the following equation:
\begin{equation} \label{eq:NS}
 \left( \begin{array}{c}
x'_{NS}\\
y'_{NS}\\
\end{array} \right) = \left( \begin{array}{cc}
r \cos \theta & -r \sin\theta \\
r \sin \theta & r \cos \theta \\
\end{array} 
\right)  \left( \begin{array}{c}
x_{NS}\\
y_{NS}\\
\end{array} \right) +
 \left( \begin{array}{c}
t_{RA}\\
t_{Decl}\\
\end{array} \right).
\end{equation}

 Calculating the transformation gives a rotation angle $\theta$ of $-0.061(31)^\circ$, $0.076(28)^\circ$, 
 $-0.018(27)^\circ$ and $0.000(29)^\circ$ and a scale factor $r$ of 1.00059(60), 1.00182(52), 1.00044(40) 
 and  1.00033(45) for the epochs 1999.97 (HRC-I), 2001.07 (HRC-S), 2005.31 (HRC-I) and 2010.61 (HRC-I),  
 respectively (numbers in parentheses represent the uncertainty in the final digits).  The values 
 of $r$ and $\theta$ for the HRC-I observations match within the 1$\sigma$ error and are significantly 
 smaller than these for the HRC-S observation. $t_{RA}$ and $t_{Decl}$ used in the translations of 
 the position of \rx~from the image- to the world coordinate system are all below $0.5''$. Indeed, 
 the largest shift is $0.29''$ for the y-coordinate in the 2010 HRC-I observation. The positions 
 of the neutron star in the four epochs are listed in Table \ref{table4}. 

 To estimate the error in the coordinates of \rx, we used the Gaussian elimination algorithm to 
 solve equation \ref{eq:transf} for $t_x$, $t_y$, $r$ and $\theta$. We then inserted these parameters 
 into equation \ref{eq:NS}. This results in equations for $x'_{NS}$ and $y'_{NS}$ that depend only 
 on values with known errors: $x_A$, $y_A$, $x_B$, $y_B$, $x'_A$, $y'_A$, $x'_B$, $y'_B$, $x_{NS}$ 
 and $y_{NS}$. The uncertainties in these two neutron star coordinates at each epoch can then be derived 
 through straightforward error propagation:  
 \begin{equation}
 \sigma _{x'_{NS} }  = \sqrt{\left( {\frac{{\partial x'_{NS} }}{{\partial x_A }}}
 \right)^2 \sigma _{x_A }^2  + 
 \left( {\frac{{\partial x'_{NS} }}{{\partial y_A }}} \right)^2 \sigma _{y_A
 }^2  +  \cdots  + 
 \left( {\frac{{\partial x'_{NS} }}{{\partial y_{NS} }}} \right)^2 \sigma
 _{y_{NS} }^2} 
\end{equation}

 The same formula is applicable for 
 $\sigma_{y'_{NS}}$. The corresponding values are listed in parentheses in Table \ref{tab:cco}.

 To check the robustness of our results we applied several cross-checks. We first repeated the 
 transformation using the fiducial points B \& C rather than A \& B. The positions of \rx\ obtained 
 this way are also listed in Table \ref{table4} for comparison. As can be seen, they have larger 
 errors than using the reference stars A \& B (because star C has only a few counts at each epoch) 
 but match the other positions within the $1\sigma$ uncertainty range. Using the combination of 
 stars A \& C rather than A \& B leads to large errors, as  A and C are located quite close 
 to one another and are in approximately the same direction relative to \rx.   In a third test 
 we calculated the position of \rx\ by applying only a two-dimensional translation of the four 
 images. We weighted the shifts of the three reference stars inversely as the variance and 
 calculated their mean for every epoch. The results for the position of \rx~differ for the 
 HRC-I observations by at most 0.4 pixel from the ones calculated according to equation 
 \ref{eq:transf}. For the HRC-S image the difference in x is $\approx 1$ pixel, though this 
 is mainly due to systematic offsets between the HRC-S and HRC-I detectors. This is also seen 
 if we compare the scale factors and rotation angles which we computed for the HRC-I and HRC-S 
 observations.

\subsection{The Proper Motion of \rx}

 To measure the proper motion of \rx~over a baseline of 3886 days we used all four positions 
 obtained from the observations between 1999.97 and  2010.61 and fitted a linear function to 
 $x'_{NS}(T)$ and $y'_{NS}(T)$ separately: 
\begin{equation}
 x'_{NS}(T)=\mu_x T + const_x,\\
\end{equation}
\begin{equation}
 y'_{NS}(T)=\mu_y T + const_y.
\end{equation}

 In these fits the projected proper motion coordinates $\mu_x$ and $\mu_y$ were taken as
 free parameters for which we find $\mu_{{\rm RA}}=-64 \pm 12\masy$  and $\mu_{{\rm Decl}}= -31 
 \pm 13 \masy$, implying a total proper motion of $\mu_{{\rm Tot}}=71 \pm 12 \masy$\@. For the 
 position angle we find $244 \pm 11$ degrees, in excellent agreement with the position angle 
 of 243 degrees as implied by the location of the optical expansion center \citep{Winkler88}. 
 The corresponding transverse space velocity for \rx\ is $(672 \pm 115) d_2 \kms$, scaled 
 relative to a distance of $d_2=d/2$ kpc. 

\begin{figure}
\centerline{\psfig{figure=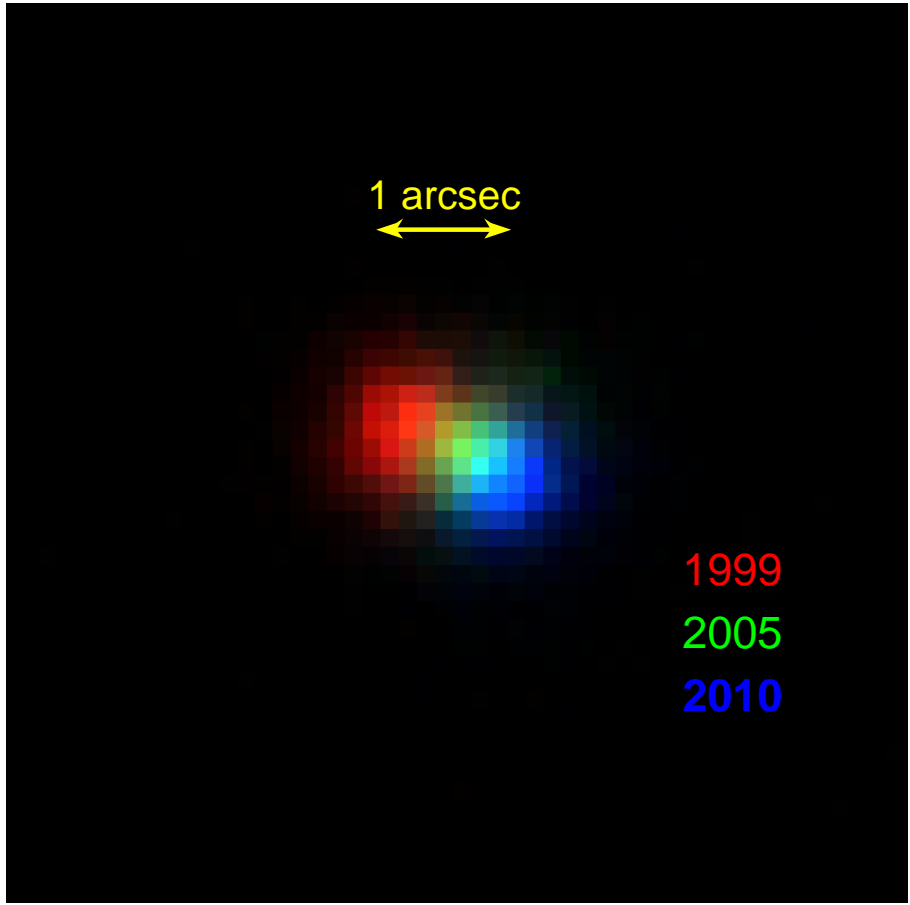,width=12cm,clip=}}
\caption[]{This enlargement of the immediate region of \rx\ shows the data from all 
three HRC-I epochs (after alignment to a common coordinate system) in different colors.   
The neutron star's motion is apparent. }
\label{figure2}
\end{figure}

\begin{figure}
\centerline{\psfig{figure=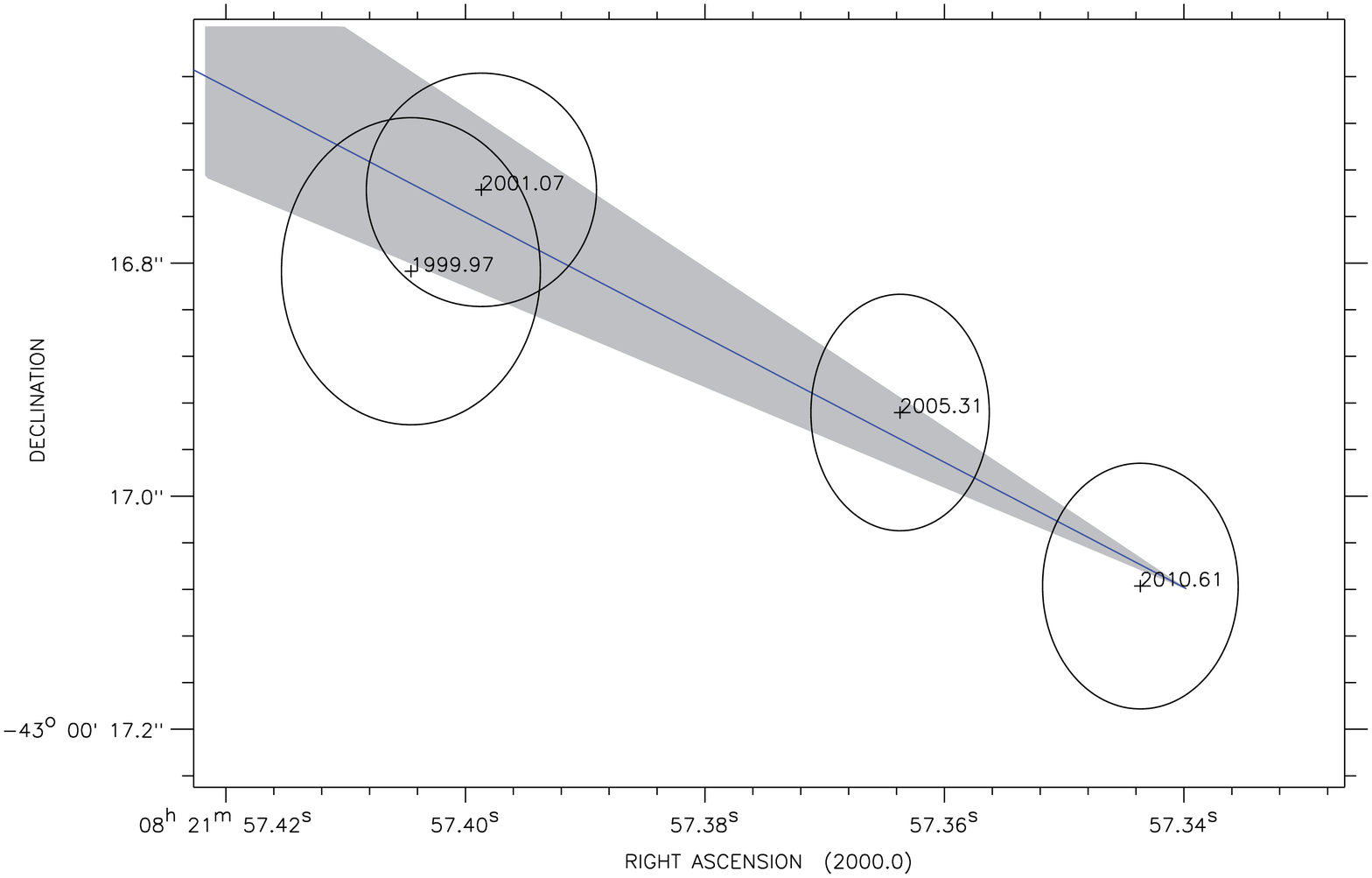,width=16cm,clip=}}
\caption[]{Position of \rx~as measured at four epochs over a baseline of 3886 days, showing 
the proper motion from upper left (NE) to lower right (SW). The $1\sigma$ position uncertainty 
at each epoch is indicated, and the observation dates are labeled. The gray 
shaded bar depicts the direction toward the remnant's optical expansion center, i.e. toward 
the birthplace of \rx. The straight blue line indicates the CCO's proper motion path as fitted 
from the four positions. 
}
\label{figure3}
\end{figure}

 Figure \ref{figure2} shows the actual data for \rx\ at the three HRC-I epochs, after alignment 
 to a common coordinate system, and Figure \ref{figure3} shows the progression of \rx\ and the 
 linear fit to its positions as measured by {\em Chandra}.

\section{Discussion}

\subsection{Kinematics and Age}
 Even though \rx\ is not moving as fast as previously thought, its association with a supernova 
 remnant of well-constrained age and reasonably well-localized explosion center \citep{Winkler88}  
 makes it worthwhile to reexamine the implications of the revised velocity on explosion kinematics 
 and dynamics. The reduced space velocity of the neutron star means a proportionately smaller 
 momentum.  Assuming a typical neutron star mass of $1.4 M_\sun$, its momentum is $\sim 2 \times 
 10^{41}\, {\rm g\,cm\,s^{-1}}$,  roughly a factor of 2 smaller than the estimate in WP07.  That
 paper compared the momentum of the neutron star with that of the optical filaments moving in the 
 opposite direction at approximately $1500 \kms$.   Assuming a typical knot mass of $0.04 M_\sun$, 
 the momentum per knot in the direction opposite the neutron star is $\sim 1.2 \times 10^{41}\, 
 {\rm g\,cm\,s^{-1}}$.  The momentum of the neutron star would thus be balanced by about 16 such 
 knots, which roughly corresponds to the number known.  Thus, to the degree of accuracy to which 
 the filament mass can be inferred, a momentum balance is thus still feasible.  

 More drastic is the reduction of the estimate of the kinetic energy carried by the stellar 
 remnant: $6.5 \times 10^{48}\, {\rm ergs}$, only a quarter of that estimated in WP07.  This 
 corresponds to only 0.6 percent of the nominal $10^{51}$ ergs produced in a canonical core-collapse 
 supernova explosion.  This reduced value undoubtedly poses less of a challenge for supernova models. 
 In fact, a potential discrepancy pointed out in WP07 is resolved.  The explosion model of 
 \citet{burrows07} predicts a kick velocity 
\begin{equation}
{V_k} \sim 1000\left( {\frac{E}{{{{10}^{51}}{\rm{ ergs}}}}} \right)\sin \alpha \kms,
\end{equation}
where $E$ represents the explosion energy, and sin$\alpha$ a parameterization of the explosion 
 anisotropy.  For the previous tangential velocity value, even for the most extreme values of 
 anisotropy, corresponding to sin$\alpha \sim 1$, the explosion energy needed to be higher than 
 the canonical $10^{51}$ ergs.  With the new, reduced tangential velocity, a range of values of 
 sin$\alpha$ = 0.6 to 0.8 is consistent with the explosion producing Puppis A having an energy 
 of $10^{51}$ ergs.  Interestingly, this range of sin$\alpha$ indicates a highly asymmetric 
 explosion.

 The revised proper motion also has mild implications regarding the age of Puppis A\@.  The proper 
 motions of the optical filaments point back to a common location at RA (2000.) = ${\rm 08^h22^m27\fs5}$; 
 Decl (2000.)$ = -42\degr57\arcmin29\arcsec$; presumably the site of the explosion (Winkler et al. 1988).  
 If no deceleration is assumed, then the remnant age inferred from the filament motion is $3700 \pm300$ years.    
 The path of the neutron star, projected backward in time, passes directly through the elliptical region 
 in which the explosion center is likely to have occurred, with a distance of $371 \pm 31$ arc seconds 
 between the current neutron star location and the nominal explosion center.  Using the new proper motion 
 measurement of $71 \pm 12 \masy$, we find that if the neutron star was born within the optically-defined 
 explosion location, its age should be $5200 \pm 1000$ years. The age inferred from the neutron star and 
 filament motions can be considered as two independent measurements of the same quantity. They average to 
 $4450 \pm 750$ years for the age of Puppis A, slightly older than the age of 3700 years based on the optical proper motions alone.

\subsection{Implications for the Kick Mechanism}
 While the velocity of \rx\ is not as extreme as previously thought, it is still high, though 
 more consistent with other fast moving neutron stars \citep[e.g.,][]{hobbs05}.  In light of 
 this new proper motion measurement we briefly reconsider the implications for kick mechanisms.  
 As discussed in WP07, electromagnetically driven or neutrino/magnetic field driven mechanisms 
 generally are not capable of providing a kick velocity in excess of $500 \kms$.  Even with the 
 lower velocity, these mechanisms are thus unlikely to be applicable to \rx.  Hydrodynamic recoil 
 mechanisms are the only ones capable of providing a sufficient kick.  In fact, recent 2D and 3D simulations 
 \citep[e.g.,][]{scheck04, scheck06, wong10, nordhaus12} suggest that kick velocities higher than $\sim 500 \kms$ are easily 
 induced by hydrodynamically driven mechanisms after about 1 second of post-bounce evolution 
 during core collapse. In these hydrodynamical explosion models the gravitational pull of slow-moving ejecta 
in front of the neutron star can -- during the first seconds of the explosion -- continue 
to accelerate it to considerably higher velocities."

 Strong independent evidence supporting a hydrodynamic kick mechanism for RX J0822--4300 comes from the 
 implications associated with the discovery and the modeling of its pulsations. \citet{gotthelf09}
 discovered a period of 112 ms in \rx.  This periodicity had been missed by prior investigators because 
 of an abrupt 180-degree phase change of the pulse profile at about 1.2 keV.  The period and the very 
 small period derivative imply a surface magnetic field strength of $\lesssim 9.8\times 10^{11}\,$G and 
 a spindown age of larger than $2.2 \times 10^5$ yr.  The very low magnetic field gives rise to the 
 attribution of ``antimagnetar" to this object.  Energy- and pulse-phase resolved spectroscopy led 
 to a model that reproduces the observed properties.  The model consists of two antipodal hotspots,
 modeled by blackbodies whose crossover in flux corresponds to the energy at which the phase reverses.  
 The angles from the hot spot to the pole and from the line of sight to the rotation axis are degenerate;  
 one has a value of  of 86\degr, the other a value of  6\degr\  \citep{gotthelf10}.  Interestingly, 
 the model degeneracy allows either for the spin axis to be parallel with the direction of motion, or 
 for the spin and kick directions to be maximally misaligned.

 The two most constraining aspects of these results on possible kick models are the low magnetic 
 field and the initial spin period.  Models invoking magnetic processes require the nascent neutron
 star to have a high magnetic field and/or high initial spin periods \citep{lai01}.  As pointed out 
 in \citet{gotthelf09}, the only current models satisfying these new constraints are the hydrodynamical 
 ones that potentially produce very high velocities.  Thus despite the less restrictive, lower inferred 
 space velocity of \rx, the conclusion of WP07 regarding likely ejection mechanisms still holds.

 The solution that favors spin-kick alignment is supported circumstantially by the fact that such 
 alignment is observed in a number of neutron stars \citep[e.g., Figure 5 in][]{wang07}.  More specific 
 support from within Puppis A comes from the alignment of HI cavities, which presumably represent the 
 path of jets from the initial explosion, along the direction of motion of the neutron star \citep{reynoso03}.  
 Further, comparison of numerical simulations with the distribution of pulsars with known spin-kick angles 
 suggests that the observed spin-kick distribution requires the initial spin period to be shorter than the 
 kick timescale \citep{wang07}.  Since the hydrodynamic recoil time scale ($\gtrsim1$ s) is considerably 
 longer than the 112 ms pulsation period, one might reasonably expect spin-kick alignment in \rx.

\section{Summary}
 Two previous proper motion measurements of \rx, incorporating the same {\em Chandra} HRC data sets, 
 produced discrepant results. HB06 found a proper motion of $104\pm 35 \masy$ at a position angle 
 of $240\degr \pm 28\degr$.  WP07, using a different analysis approach, found a considerably 
 higher value of $165\pm 25 \masy$ at a position angle of $248\degr \pm 14\degr$.  A combined 
 analysis, incorporating a new deep observation and a total time baseline twice as long has 
 was available previously, yields an improved value that is smaller than either of the two 
 published numbers, but at a position angle consistent with both. We now find with high 
 confidence a value for the proper motion of $71 \pm 12 \masy$ at a position angle of 
 $244\degr \pm 11\degr$.  For an assumed distance to Puppis A of 2 kpc, the proper motion 
 corresponds to a tangential velocity of $672 \pm 115 \kms$.  This smaller velocity eases 
 most challenges that \rx\ previously posed to pulsar-kick models, but still requires a 
 hydrodynamic kick model in a highly asymmetric explosion. Considering the proper motion
 based age estimate as a second and independent measurement to the one inferred from the 
 motion of optical filaments, it can be averaged to $4450 \pm 750$~yrs which implies that
 Puppis A is slightly older than the  3700 years based on the proper motions of the optical filaments.

\acknowledgments
\noindent{\bf Acknowledgments}\linebreak
 TP acknowledges support by the International Max-Planck Research School on 
 Astrophysics at the Ludwig-Maximilians University, IMPRS; PFW acknowledges 
 support from the NSF through grant AST-0908566.  We are grateful to Daniel
 Patnaude for his contributions in scheduling and setting up the observations 
 and to several members of the Helpdesk staff at the {\em Chandra} X-ray Center 
 for assistance at various stages of this project, and especially for helping 
 to sort out the troublesome software bug described in \S2.1. We also acknowledge 
 the use of the {\em Chandra} data archive.

\clearpage

\clearpage

\clearpage

\clearpage

\begin{deluxetable}{ccccc}
\tabletypesize{\small}
\tablewidth{0pc}
\tablecaption{{\em Chandra} observations of the neutron star in \pu \label{table1}}
\tablehead{
 Instrument &   ObsId &    Date (MJD)           & ONTIME   & Exposure Time    \\ 
     { }    &    {}   &        {}               &  (sec)   &     (sec)          }
\startdata        
   HRC-I    &    749  &  21/22 Dec 1999 (51533) & 18014  &    $\,\,\;$9907        \\ 
   HRC-S    &   1859  &  25 Jan 2001 (51934)    & 19667  &    19524        \\ 
   HRC-I    &   4612  &  25 Apr 2005 (53485)    & 40165  &    21410        \\  
   HRC-I    &   11819 &  10/11 Aug 2010 (55418) & 33681  &    15509        \\
   HRC-I    &   12201 &  11 Aug 2010 (55419)    & 38680  &    17855        \\
\enddata                                                                                                                     
\end{deluxetable}                                                                                                   

\clearpage

\begin{deluxetable}{ccccccc}  
\tabletypesize{\small}
\tablewidth{0pc}
\tablecaption{Position, proper motion and angular distance of astrometric reference 
stars near  \rx~as listed in the UCAC3 catalog \citep{Zacharias09}. \label{table2}}
\tablehead{
\multicolumn{2}{c}{Designation} &\multicolumn{2}{c}{Position (J2000.0)} & \multicolumn{2}{c}{Proper Motion } &
$\delta d$  \\
 Short   & 3UCAC         & R.A.          & Decl.         & $\mu_{R.A.}$  & $\mu_{Decl.}$ &  \\
         &               & (h m s)       & (d m s)     & mas yr$^{-1}$ & mas yr$^{-1}$ & arcmin }
\startdata
  A & 094-058669  &08 21 46.292(1)&-43 02 03.64(5)& $-14.3\pm 2.0$ &$-3.6\pm 5.5 $ & 2.8 \\
  B & 095-060051  &08 22 24.003(3)&-42 57 59.37(2)& $0.0  \pm 4.0$ &$10.2\pm 2.0 $ & 5.3 \\
  C & 094-058675  &08 21 48.876(6)&-43 01 28.33(6)& $-55.4\pm 8.6$ &$2.0\pm 6.1 $ & 2.0 \\ 
\enddata  
\tablecomments{$1\sigma$ uncertainty of the last digit in parentheses}
\end{deluxetable} 

\clearpage  

\begin{deluxetable}{ccccccrcc}
\tabletypesize{\scriptsize} 
\tablewidth{0pc}
\tablecaption{Properties of the astrometric reference stars
\label{table3}\label{tab:stars2}}
\tablehead{ {} & {} &  {}  &\multicolumn{2}{c}{X-ray$^a$}  & & &  \multicolumn{2}{c}{Optical$^a$}  \\
 ObsID       & Epoch  & Source & R.A.          & Decl.            & Counts & Rate$\,\,$ & R.A. & Decl.        \\
   {}        &        &        & (h m s)       & (d m s)        &         & cts/ks     & (h m s)  & (d m s)  }
\startdata                                                                          
   749       &1999.97 &  NS  & 08 21 57.411(01) & -43 00 16.63(01) & 2544 & 257.0 &      {}          &           {}        \\
             &        &   A  & 08 21 46.295(10) & -43 02 03.26(17) &   46 &   4.6 & 08 21 46.292(01) & -43 02 03.64(02)    \\
             &        &   B  & 08 22 24.008(25) & -42 57 59.58(17) &   46 &   4.7 & 08 22 24.003(03) & -42 57 59.37(04)    \\
             &        &   C  & 08 21 48.874(18) & -43 01 28.13(26) &   13 &   1.3 & 08 21 48.876(06) & -43 01 28.33(09)    \\

  1851       &2001.07 &  NS  & 08 21 57.390(01) & -43 00 16.91(01) & 5005 & 256.5 &      {}          &           {}        \\
             &        &   A  & 08 21 46.316(09) & -43 02 03.77(11) &   70 & 3.6 & 08 21 46.291(01)   & -43 02 03.64(02)    \\
             &        &   B  & 08 22 23.930(21) & -42 57 59.38(20) &  139 & 7.1 & 08 22 24.003(03)   & -42 57 59.36(04)    \\
             &        &   C  & 08 21 48.880(17) & -43 01 28.62(28) &   11 & 0.5 & 08 21 48.870(07)   &  -43 01 28.33(09)   \\

   4612      &2005.31 &  NS  & 08 21 57.374(01) & -43 00 17.07(01) & 5596 & 261.6 &      {}          &           {}        \\
             &        &   A  & 08 21 46.297(08) & -43 02 03.70(08) &   91 &   4.3 & 08 21 46.285(02) & -43 02 03.66(05)    \\
             &        &   B  & 08 22 24.006(11) & -42 57 59.60(23) &   57 &   2.7 & 08 22 24.003(04) & -42 57 59.31(05)    \\
             &        &   C  & 08 21 48.881(22) & -43 01 28.40(26) &    7  &  0.3 & 08 21 48.849(10) & -43 01 28.32(11)    \\

11819/12201  &2010.61 &  NS  & 08 21 57.329(01) & -43 00 17.38(01) & 9152 & 274.7 &      {}          &           {}        \\
             &        &   A  & 08 21 46.268(08) & -43 02 03.93(07) &  132 &   4.0 & 08 21 46.279(03) & -43 02 03.68(08)    \\
             &        &   B  & 08 22 23.979(13) & -42 57 59.60(23) &   96 &   2.9 & 08 22 24.003(07) & -42 57 59.26(06)    \\
             &        &   C  & 08 21 48.853(22) & -43 01 28.68(17) &   21 &   0.6 & 08 21 48.822(14) & -43 01 28.31(15)    \\
\enddata  
\tablecomments{$^a 1\,\sigma$ uncertainty of the last two digits in parentheses}
\end{deluxetable}

\clearpage

\begin{deluxetable}{ccccccc}
\tabletypesize{\small}
\tablewidth{0pc}
\tablecaption{Positions of \rx \label{table4} \label{tab:cco}}
\tablehead{
  Epoch &  Ref.~Stars   & R.A.             & Decl.                \\
    {}  &     {}        & (h m s)          & (d m s)            }
\startdata
1999.97 & A \& B        & 08 21 57.403(11) & -43 00 16.80(13)   \\
        & B \& C        & 08 21 57.409(16) & -43 00 16.72(21)   \\
2001.07 & A \& B        & 08 21 57.398(10) & -43 00 16.74(10)   \\
        & B \& C        & 08 21 57.403(15) & -43 00 16.65(22)   \\
2005.31 & A \& B        & 08 21 57.363(08) & -43 00 16.93(10)   \\
        & B \& C        & 08 21 57.349(19) & -43 00 16.92(22)   \\
2010.61 & A \& B        & 08 21 57.343(08) & -43 00 17.08(11)   \\
        & B \& C        & 08 21 57.312(21) & -43 00 16.99(18)   \\
\enddata  
\tablecomments{$1\sigma$ position uncertainty in parentheses}
\end{deluxetable}

\end{document}